\documentclass[aps,prb,superscriptaddress,twocolumn]{revtex4-2}
\usepackage{amsmath}
\usepackage{amssymb}
\usepackage{lipsum,graphicx}
\usepackage{amsfonts}
\usepackage{color}
\usepackage{graphicx}
\usepackage{amsfonts}
\usepackage{textcomp}
\usepackage[normalem]{ulem}
\newcommand{\beq}{\begin{equation}}
\newcommand{\eeq}{\end{equation}}

\usepackage{hyperref}

\begin{document}
\title{Ferromagnetic resonance in an antiferromagnetic crystal EuSn$_2$As$_2$}
\author{I. I. Gimazov}
\author{D. E. Zhelezniakova}
\author{R. B. Zaripov} 
	\affiliation{Zavoisky Physical-Technical Institute, FRC Kazan Scientific Center of RAS,  Kazan 420029, Russia}
\author{Yu. I. Talanov}
\email[E-mail:]{talanov@kfti.knc.ru}
	\affiliation{Zavoisky Physical-Technical Institute, FRC Kazan Scientific Center of RAS,  Kazan 420029, Russia}
\author{A. Yu. Levakhova}
\author{A. V. Sadakov}
\author{K. S. Pervakov}
\author{V. A. Vlasenko}
\affiliation{V. L. Ginzburg Research Center, P.N. Lebedev Physical Institute RAS, Moscow 119991, Russia}
\author{A. L. Vasiliev}
\affiliation{National Research Center``Kurchatov Institute'', Moscow 123182, Russia}
\affiliation{Moscow Institute of Physics and Technology, Dolgoprudny, Moscow district, 141701, Russia}
\author{V. M. Pudalov}
\affiliation{V. L. Ginzburg Research Center, P.N. Lebedev Physical Institute RAS, Moscow 119991, Russia}

\begin{abstract}
We report results of  electron spin resonance (ESR) measurements
in single crystals of EuSn$_2$As$_2$. In the temperature range of antiferromagnetic (AFM) ordering of Eu atoms, $T \leq T_N\approx 24$\,K, the ESR signal splits into two resonance lines, 
one of which, at high-field (or low-frequency), is the conventional acoustic AFM resonance mode that occurs at temperatures below $T_N$.
The lower-field (high-frequency) line, as we  have proven here,  is the ferromagnetic resonance associated with the presence  in the layered AFM crystal of a small amount ($\sim 3\%$) of planar nanodefects with a non-zero ferromagnetic (FM) moment.  
The existence of ferromagnetic nano-inclusions in the bulk of the antiferromagnetic compound makes EuSn$_2$As$_2$ a peculiar example of a natural magnetic metamaterial.  We believe that the
planar FM nanodefects are also inherent in other layered AFM compounds, which explains often observed increase in their magnetic susceptibility upon cooling at $T< T_N\rightarrow 0$. 
\end{abstract}
\maketitle

\section{Introduction}
Time reversal symmetry plays a critical role in the quantum properties of materials.
To control it, magnetic impurities are usually introduced into the materials being studied, for example, Mn in GaAs.
Doping, however, leads to strong disorder of the crystal,
masking the quantum effects under study. Stoichiometric pure compounds with antiferromagnetically (AFM) ordered atoms in the crystal lattice allow one to avoid disorder.
In this regard, layered van der Waals pnictides and chalcogenides with antiferromagnetic (AFM) ordering of magnetic atoms (Eu, Mn, Co, etc.) provide a successful platform for such studies. 
This class of AFM stoichiometric compounds 
includes topological insulators (e.g., MnBi$_2$Te$_4$), Weyl semimetals (EuCd$_2$As$_2$), axion insulators (EuIn$_2$As$_2$), as well as materials with a simpler spectrum - non-Weyl semimetals EuFe$_2$As$_2$, EuRbFe$_4$As$_4$, EuSn$_2$As$_2$, EuSn$_2$P$_2$, CsCo$_2$Se$_2$, CaCo$_2$As$_2$, etc.
All these compounds attract a great research interest  for their non-trivial properties, determined by the mutual interplay of the topology of their band structure,  physics of  magnetic ordering and superconducting pairing,  as well as exchange and spin-orbit interactions 
\cite{otrokov_Nature_2019, li_PRX_2019, golov-EuFeAs2_PRB_2022, EuRb1144_PRB, jo-EuCd2As2_PRB_2020, yu-EuIn2As2_PRB_2020, dietl_PRL_2023, xu-In2As2_PRL_2019, pakhira-EuSn2As2_PRB_2021, lv_ASCAplElextronMat_2022, CommunMat_2025, PRB_tbp}. 

  This work focuses on  a layered semimetallic compound of the stoichiometric composition EuSn$_2$As$_2$ (ESA), in which Eu atoms are ordered as the temperature decreases below $T_N \approx 24$\,K, forming an A-type antiferromagnet structure  with an easy  $ab$ plane \cite{pakhira-EuSn2As2_PRB_2021}.
  Despite the apparent simple magnetic arrangement, the high-purity single crystals of this  compound, in the AFM state, exhibit nontrivial properties inconsistent with those of the conventional single-phase antiferromagnet. 
  In particular, in  a weak DC magnetic field, 
 the AC magnetic susceptibility tends to increase below $T_N$\ or even to diverge as $T\rightarrow 0$ \cite{chen_ChPhysLett_2020, li_PRB_2021, li_PRX_2019, ESA-defects, pakhira-EuSn2As2_PRB_2021}, 
 the  DC magnetization dependence on external field applied in the $ab$ plane exhibits a weak nonlinearity \cite{golov_JMMM_2022}  and even a hysteresis \cite{ESA-defects}; these effects are pronounced in low fields $H\ll H_{sf}$, much less than the field of complete spin polarization $H_{sf} \approx 4$T \cite{golov_JMMM_2022}.
 
 In addition, earlier in Ref.~\cite{golov_JMMM_2022} we observed that the electron spin resonance (ESR)  line in the AFM state is split into two components which have significantly  different magnetic field  dependences. 
 While the main high-field (low-frequency) ESR line (the so-called ``A-line'') \cite{golov_JMMM_2022} is 
 identified as the acoustic mode of spin vibrations, the origin of the lower-field (high-frequency) ``M-line'' remained a mystery until to date; in \cite{golov_JMMM_2022} it was only shown that the M-line 
 can not be associated with the optical mode of antiferromagnetic resonance. Potentially, these puzzling features 
 might be related with our recent finding in Ref.~\cite{ESA-defects} where by 
 high angle annular dark field scanning transmission electron microscopy (HAADF STEM)
 we revealed the presence of planar defects in ESA single crystal and by DFT calculations have shown they to be the source of local ferromagnetism.

In this work, we carried out more detailed ESR measurements, with higher sensitivity (achieved through the use of a microwave cavity spectrometer), and over a wide temperature range, from $T=4{\rm K\,} (\ll T_N)$ to $300{\rm K\,} (\gg T_N)$. As a key for interpreting the ESR data and following the results of 
Ref.~\cite{ESA-defects}, we apply a model of a metamaterial, with ferromagnetic nano-stripes embedded into the antiferromagnetic crystal.  The novel and previous ESR data measured with ESA crystals 
are shown to be consistent with such simple model of the FM/AFM metamaterial. We conclude that  the anomalous resonant M-line indeed  has a ferromagnetic origin, and is associated with nano-scale planar defects present in ESA single crystals. 

\subsection{Samples studied}
\subsubsection{Crystalline and magnetic structure}
\begin{figure}[h]
		\includegraphics[width=230pt]{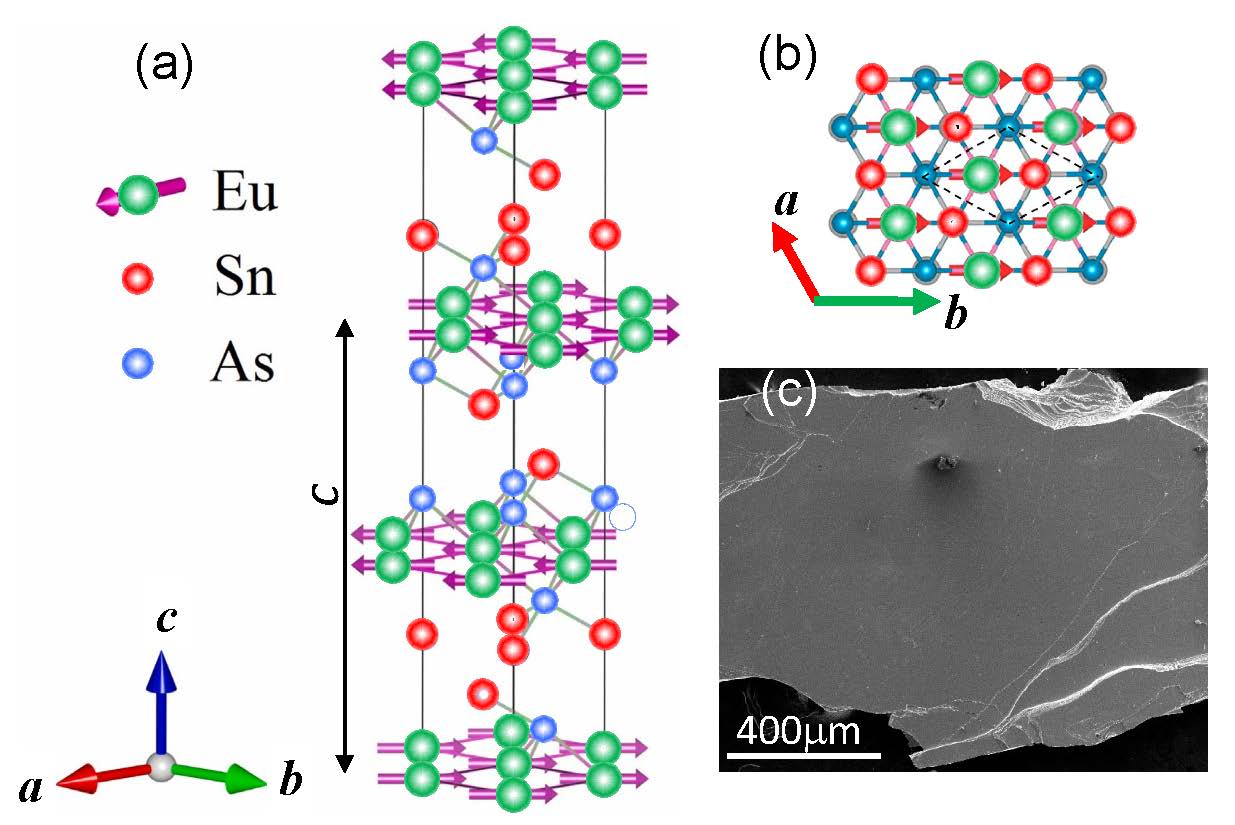}
	\caption{(a,b) Crystal lattice structure of EuSn$_2$As$_2$ in the $ac$- and $ab$- planes, respectively. Crimson arrows depict magnetic moments 	of Eu atoms in the AFM state 	(adapted from \cite{golov_JMMM_2022}), 
		(c) SEM image of a typical crystal (field of view $1300 \mu$m wide). 
		}
	\label{fig:crystals}
\end{figure}

EuSn$_2$As$_2$ crystallizes into a Bi$_2$Te$_3$ lattice of the trigonal symmetry group $R\overline{3}m$. 
The crystals are layered with van der Waals bonding between the Eu--SnAs--SnAs--Eu layers. The seven-layer unit cell 
  has a large size along the $c$ axis  ($c=26.4$\AA) and consists of four SnAs bilayers located between the three  Eu-layers (see Fig.~\ref{fig:crystals}a ). 
In the $ab$ plane, the Eu atoms form a triangular lattice 
 (hexagonal unit cell) with each Eu atom coordinated
by six As atoms forming a trigonal prism (Fig.~\ref{fig:crystals}b).
Eu atoms in individual layers are ferromagnetically polarized.  At temperatures below 24\,K the magnetic moments of Eu atomic layers undergo an  A-type antiferromagnetic ordering, in which the magnetization vector of Eu layers rotates by $\pi$ from layer to layer, remaining in the same $ab$ plane. This magnetic structure has been determined from neutron diffraction measurements \cite{pakhira-EuSn2As2_PRB_2021} and also indirectly follows, e.g.  from  anisotropic magnetization measured in DC fields \cite{golov_JMMM_2022, CommunMat_2025}. In the measurements, at $T\ll T_N$ the magnetization follows about a liner field dependence and sharply saturates at $H_s\approx 4$\,T,	where the anisotropic $H_s$ value  corresponds to complete spin polarization  and confirms the easy-plane AFM state \cite{golov_JMMM_2022, PRB_tbp}.

\subsubsection{Crystal synthesis and growth}

For high-temperature synthesis (same as in Ref.~\cite{golov_JMMM_2022}), we used the high purity starting materials: 99.99\% Sn, 99.9999\% As, and Eu  (99.95\%). The precursor SnAs  was obtained through the high-temperature solid-phase synthesis reaction. Further, the  SnAs precursor and Eu chips were mixed in stoichiometric molar ratio 2:1; the mixture was then placed in  an alumina crucible in a sealed quartz ampule filled with  Ar-gas. The ampule was heated in an oven up to  850$^\circ$C,  held for 12\,h at this temperature. Then the melt was cooled to 550$^\circ$C  at a rate of 2$^\circ$C/h, and annealed during 36\,hours.  After this the oven was switched off and slowly cooled down with the ampule. 

 This synthesis protocol, namely with such stoichiometric components composition, 
according to Ref.~\cite{pakhira-EuSn2As2_PRB_2021} provides the   homogeneous phase composition \cite{note}.
The EDX analysis of the grown crystals showed a homogeneous spatial distribution of elements with average  ratios   
1\,:\,2.015\,:\,1.85 (Eu:Sn:As)  - with a minor deficiency of As. 
For measurements, we have selected  EuSn$_2$As$_2$ single crystals with size up to $3\times 2\times 0.2$mm$^3$; 
the scanning electron microscope (SEM) image in  Fig.~\ref{fig:crystals}c shows their mirror smooth surface.

\subsubsection{Bulk lattice structure, local atomic and magnetic structure}

\begin{figure}
		\includegraphics[width=240pt]{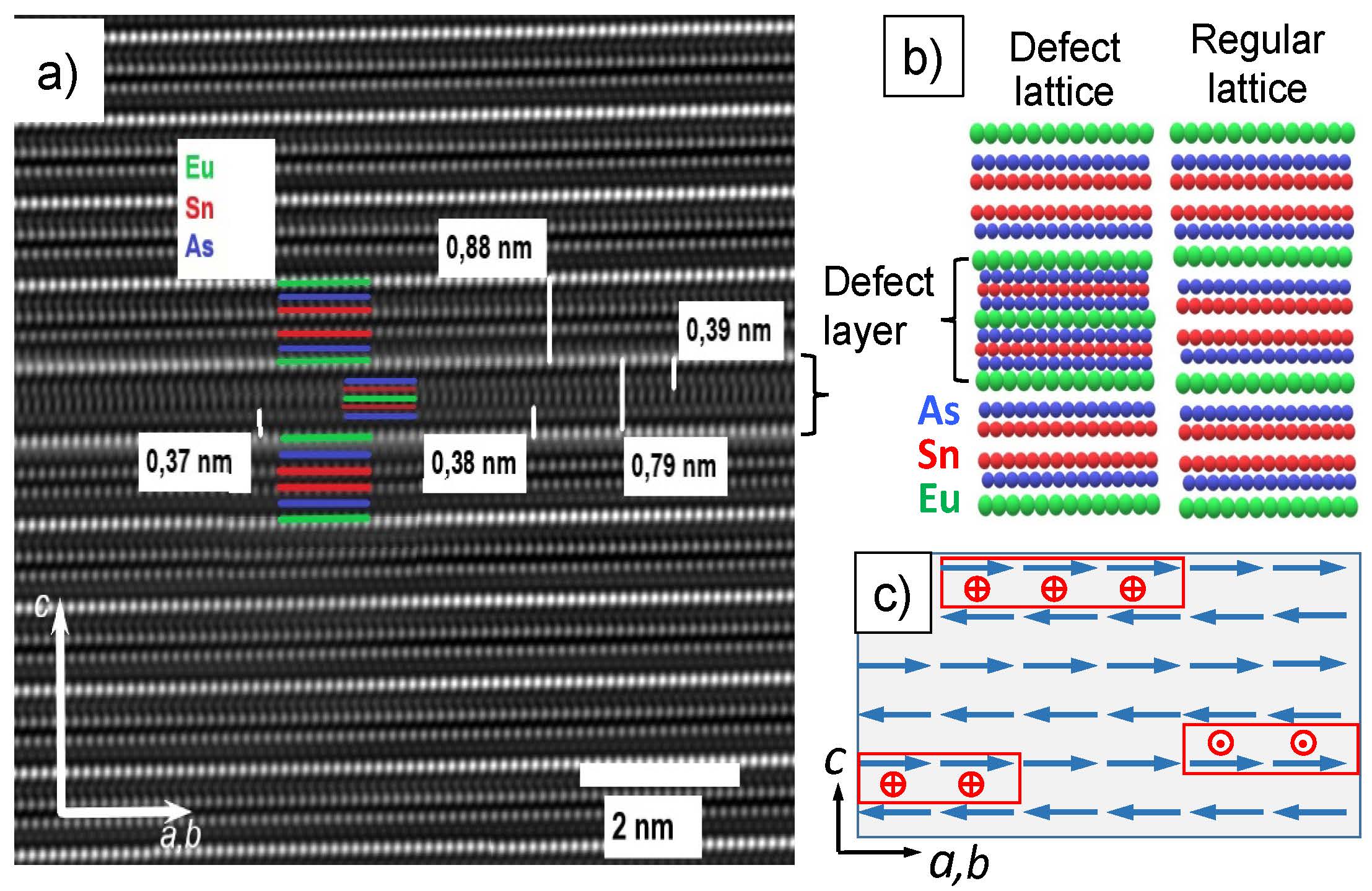}
	\caption{(a) FFT filtered  high-resolution HAADF STEM
		image of the crystal lattice cross section in the $ac$ plane (adapted from Ref.~\cite{ESA-defects}). The curly bracket marks the 0.79nm thick defect layer with almost missing one row of Sn atoms and with an extra row of Eu atoms;
		(b)Schematic arrangement of several layers of the  lattice in the $ac$ plane: regular lattice (right), and  lattice with a defect layer (left);
		(c)Schematics of the conjectural  	model of a metamaterial
		structure with planar defects and of the local magnetic ordering in the AFM state.  Red arrows show magnetic  moments of Eu atoms. 
		The planar defects are outlined by  rectangular  frames, where the blue circles with crosses and dots show possible directions of their FM-moments.
		}
	\label{fig:defects}
\end{figure}

Precise X-ray diffraction (XRD) measurements  confirmed high perfection of the crystal lattice in  the bulk of the studied crystals \cite{PRB_tbp}. 
Nevertheless, transmission electron microscopy (TEM)  of the  local lattice structure with atomic resolution   \cite{ESA-defects} has revealed a noticeable amount  (about 3\%) of nanostructural planar defects, the example is shown in Fig.~\ref{fig:defects}a. The planar defect represents 
a pair of layers of Eu atoms located closer to each other 
along the $c$-axes, 
and an intermediate Eu-layer replacing one missing Sn layer  (see Fig.~\ref{fig:defects}, and also 
Ref.~\cite{ESA-defects}).

The simplified  formula for the local chemical composition in the nanodefect area 
is  Eu$^{(2+)}$Sn$^{(4+)}$As$^{(3-)}_2$ 
 \cite{ESA-defects}, the known cubic phase \cite{database}. It differs from the bulk compound Eu$^{(2+)}$Sn$_2^{(2+)}$As$_2^{(3-)}$ due to the changed valence of Sn. 
The incorporation of the planar defect with its own cubic lattice into the  trigonal  
{\em R\={3}m} 
lattice of the ideal EuSn$_2$As$_2$ host crystal occurs though the (111)  plane of the defect, and involves  participation of an entire transition block.  We treat the latter one as an extended defect cell,  that  together with the planar defect itself, is described by the general formula Eu$_7$Sn$_{12}$As$_{14}$ and hence consists of 33 atoms, including 7 Eu atoms in the minimum unit cell of the extended defect block \cite{ESA-defects}.

As mentioned above, in the ideal lattice of the
host crystal  at $T< T_N$, the magnetic moments of  Eu atoms are aligned ferromagnetically within the layer, and compensate each other due to antiferromagnetic  interlayer A-type ordering.
In contrast, in the defect area, an odd number of Eu atoms per extended  cell (7) should inevitably lead to an uncompensated magnetic moment. This is confirmed by our DFT calculations  \cite{ESA-defects}, which showed that the defect has a local ferromagnetic moment $\sim (1/7)m_{\rm Eu}$/f.u., where $m_{\rm Eu}=6.77\mu_B$ \cite{EuRb1144_PRB, golov_JMMM_2022, golov-EuFeAs2_PRB_2022, talanov_JETPL}. 
In other words, exactly one of the 7 Eu atoms included in the extended defect formula undergoes ferromagnetic ordering, while the remaining 6 atoms have  AFM-compensated moments.

The ferromagnetism of defects is consistent with other features experimentally and theoretically identified in Ref.~\cite{ESA-defects}:  weak FM-hysteresis in the DC magnetization curves in low fields $H\ll H_s$ (less than the spin canting field $H_s \sim 0.01T$) and in the temperature range of AFM-ordering, $T< T_N$, as well as an anomalous increase in the AC susceptibility as the temperature decreases below $T_N$. 

\subsubsection{The approach used and modeling of the object under study}
The main goal of this work is to study in more detail the puzzling spectral structure of the ESR observed in the EuSn$_2$As$_2$ crystal and to test our assumption that the anomalous resonance line is related to ferromagnetic 
defects. 

According to the information on planar defects  presented in Ref.~\cite{ESA-defects}, we further 
treat the system under study as a metamaterial, where  the host matrix of AFM-ordered Eu layers contains 
planar defects. The individual defect possesses a ferromagnetic moment of 1/7 of $N_{\rm def} \mathfrak{m}_{\rm Eu}$ (where $N_{\rm def}$ is the number of Eu atoms per defect, and $\mathfrak{m}_{\rm Eu} =6.77\mu_B$. 
This model is schematically illustrated  in Fig.~\ref{fig:defects}c. 
For reasons of minimum free energy, the uncompensated ferromagnetic moments 
should lie in the $ab$ plane. This is also consistent with (i)  the observation of the ferromagnetic-type magnetization hysteresis namely in the  $H\| (ab)$ plane geometry and (ii) the absence of hysteresis for $H\perp (ab)$ plane \cite{ESA-defects}. We don't have  enough information to conclude on the direction of ferromagnetic moments in the $ab$-plane.
However, this information is insignificant for the purpose of this paper, and for simplicity  we assume 
the FM-moments are perpendicular to the AFM vector. Also, various  defects may have opposite magnetic moment directions, equally probable, as schematically shown in  Fig.~\ref{fig:defects}c. Additionally, some defects which have sufficiently large width $W$ may contain FM-domains with oppositely directed magnetization vectors (see Fig.~\ref{fig:defects}c). 

 The Eu atoms with uncompensated moment inside one defect are located close to each other, 
 their ferromagnetic moments are correlated with each other and therefore respond to the microwave field as a single ferromagnet. By contrast, individual defects are located macroscopically far apart and therefore are uncorrelated.
 Indeed, magnetic field created by the defect decays with distance $r$ as $(d,W/r)^3$, where $d\approx 0.5$nm is the thickness of the defect layer in the $c$ direction,  
 and $W\sim 100$\,nm is the characteristic width of the defect stripe along the $a$ (or $b$) axis.  
 The stray field, therefore, is negligibly small and the defects do not ``feel'' each other. 

For the stated reason, the FM moments of  defects resonate in the microwave field individually, 
and the number of FM-defects (as well as the total FM moment) in the crystal determines only the amplitude 
of resonant absorption, but not the resonance frequency.  Thus, the saturation FM-magnetization, which determines the frequency of ferromagnetic resonance, depends on the average 
magnetic moment of one defect per unit volume of the given defect $\mathfrak{m}_{\rm def}$, rather than  per unit volume of the crystal. Hence, it is reduced only by a factor of $\sim 7$  compared to the AFM saturation magnetization. In the framework of this model,  the ratio of the FM magnetization $\mathfrak{m}_{\rm def}$ {\em per unit volume of the defect} to the AFM effective magnetization {\em per unit volume of the crystal} $\mathfrak{m}_{\rm AFM}$ is 1:7.  As shown below, this estimate roughly corresponds to 
the ratio of  saturation magnetization fields 1:9.6, determined from the temperature and field dependence of the resonant frequencies/fields in the  AFM and FM resonances.

The model under consideration assumes that the sample consists of 
independent ferromagnetic strips (with the length to width ratio $L/W \gg 1$ \cite{ESA-defects}) 
located far  from each other in the AFM-matrix. 
In such composite sample, the FM resonance frequency does not depend on the number of pieces per unit volume,
until they are brought together to form a single magnet. This model explains the simultaneous existence of two resonances in the ESR spectrum: (i) AFM resonance on the bulk ``correct'' magnetic sublattice of Eu spins and (ii) FM resonance on Eu spins in the regions of FM nanodefects.

In contrast to what was said about ESR, we note that in the DC magnetization of the crystal  the ferromagnetic contributions of various defects \cite{ESA-defects} is summed over all defects and is determined by the total ({\em or average per volume}) FM magnetization   of the crystal.  
In this case, (and taking into account 3\% volume share of the defects in the studied sample) we can expect that the DC saturation magnetization should be $\sim 7\times (3\%)^{-1}\sim 230$ times less than the saturation magnetization 
of the AFM crystal. This estimate roughly corresponds to the ratio of 300:1 for the saturation values of 
ferromagnetic and antiferromagnetic magnetizations measured in Ref.~\cite{ESA-defects}.

We compare below the measured dependencies of the resonance frequencies on temperature and on magnetic field and demonstrate their agreement with our model of a metamaterial (i.e. mixed AFM/FM state). This agreement between the model (supported by the DFT calculations \cite{ESA-defects}) and the measured dependencies 
confirms that the ``puzzling'' high-frequency M-line in the ESR spectrum is a consequence of the presence of a small fraction of FM nanodefects in the volume of the host AFM crystal. 
Despite the small number of nanodefects ($\sim 3\%$), the ferromagnetic nature of this resonance significantly enhances its contribution to the total ESR signal from the entire crystal, making the amplitudes of AFM and FM resonant absorption comparable.

 \subsection{Measurements of resonant microwave absorption}

 \subsubsection{Measurement technique}

The measurements were carried out using two microwave spectrometers (Bruker BER-418s with a fixed frequency of $\sim 9.35$\,GHz and Bruker Elexsys E-580 with a frequency of $\sim 9.7$\,GHz)
while varying the magnetic field from 0 to 1.5\,T at temperatures from 4.2K to 296\,K. 
The sample was placed in a cylindrical  TE$_{011}$ cavity  at the antinode of the magnetic component of the microwave field. In all measurements, the magnetic component of the microwave field $H_\sim$ was oriented perpendicular to the direction of the constant magnetic field, as required to excite electron spin resonance. The observed resonance is associated with Eu atoms having a large spin magnetic moment $\approx 6.77 \mu_B$ ($J=7/2$) 
\cite{golov_JMMM_2022, kim_PRB_2021, stolyarov_JPCL_2020}. The use of microwave cavity resonators and magnetic 
field modulation (at 100\,kHz) with phase-sensitive detection of the absorption signal made it possible to detect even a weak resonance signal at high temperatures.

\subsubsection{Temperature dependence of the resonance signal shape} 

\begin{figure}
		\includegraphics[width=115pt]{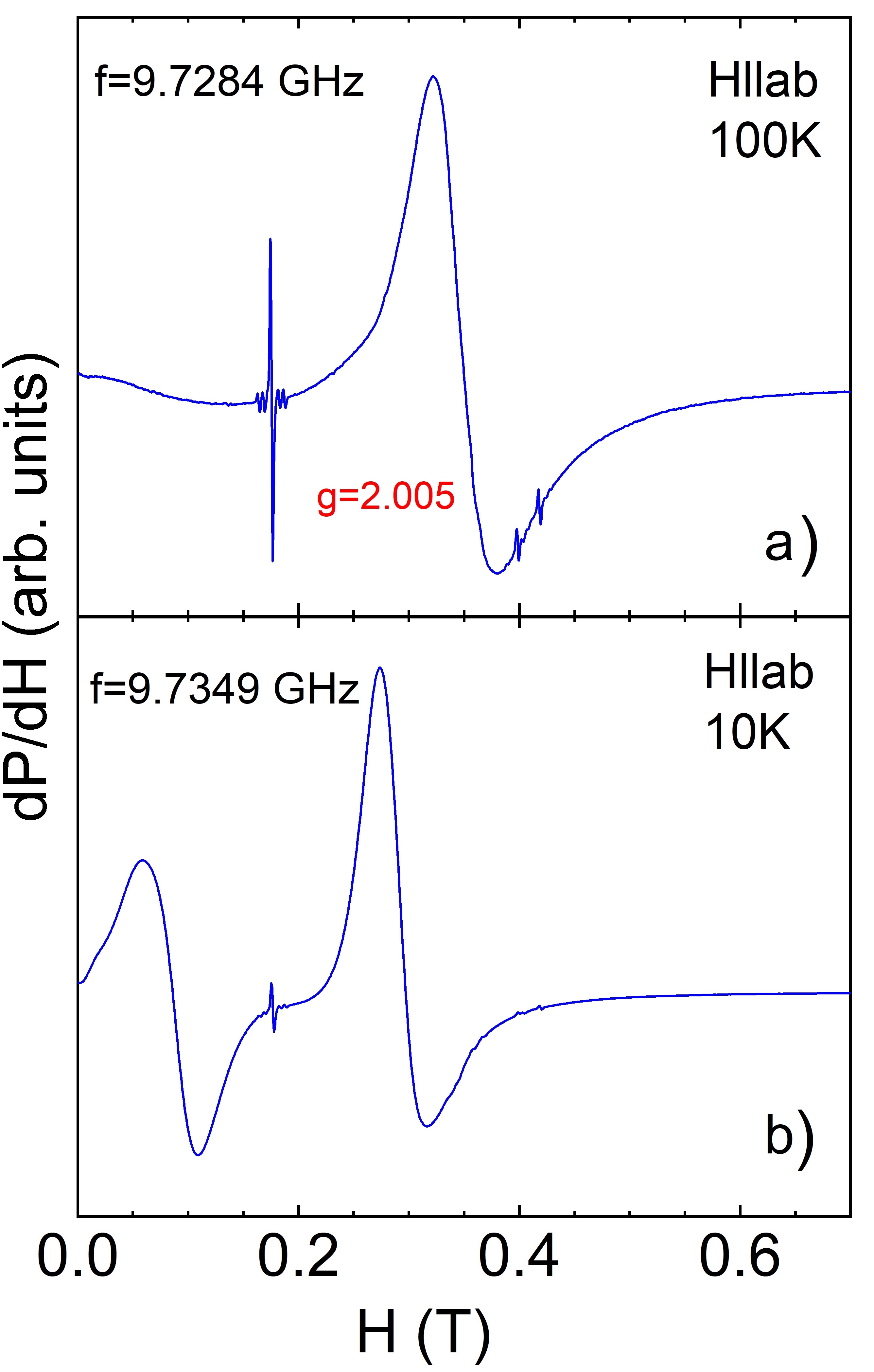}
		\includegraphics[width=114pt]{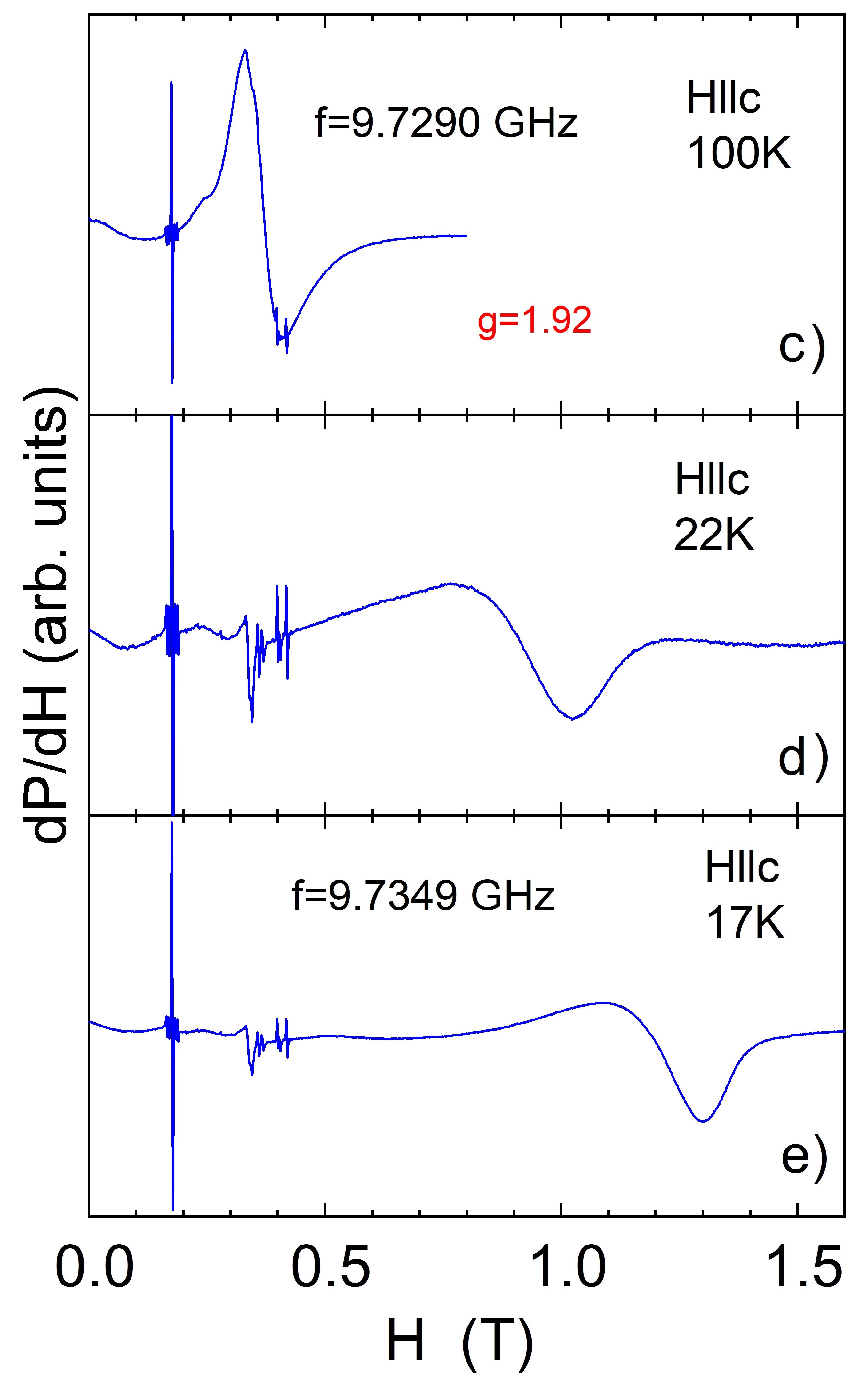}
	\caption{Typical shape of the resonance absorption signal: (a, b) - in the $H\parallel ab$ geometry at two temperatures, and (c, d, e) - in the $H\|c$ geometry at three temperatures. 
		Representative temperatures are above, near and below $T_N \approx 24$K.
The shape of the resonance signal in panels (d) and (e) has an inverse asymmetry (compared to the Dyson shape)
\cite{note:line-shape}. 	    
Other narrow features noticeable in the resonant absorption spectra in weak fields $H<0.5$T are associated with impurities in the resonator material (sapphire).}
	\label{fig:spectra}
\end{figure}

The evolution of the ESR signal shape as the temperature changes from $T=100$K (obviously, in the paramagnetic region) to temperatures below $T_N$ is shown in Figure \ref{fig:spectra} for two magnetic field orientations: 
in the easy $ab$-plane and perpendicular to it. At high temperatures ($T\gg T_N$), the resonance spectrum consists of one line, the position of which weakly depends on the orientation of the sample relative to the direction of the DC magnetic field (and corresponds to the values of the $g$-factor, $\textit{g}=2.005$ for $H\|(ab)$\ and $g=1.92$ for $H\|c$.

At low temperatures $T<T_N$ for the direction of the field in the $(ab)$ plane, the resonance is split into two lines of comparable amplitude: in addition to the usual acoustic mode of the AFM resonance (the so-called A-line)
in weaker fields a second mode (M-line) appears. The high-field signal (A-line) originating from the bulk was modeled by the Dyson function, due to the skin effect. The low-field signal (M-line), which we believe originates from thin planar defects much smaller than the thickness of the skin layer, is approximately symmetrical and is therefore approximated by the Lorentz function. 

For another field orientation, $H\|c$, i.e. perpendicular to the easy magnetization plane, the
splitting at $T<T_N$ is also present, but the amplitude of the low-field line is much weaker (see Fig.~\ref{fig:spectra}d,e). 
The splitting of the signal into two lines is visible only in the immediate vicinity of the Neel temperature $T\approx T_N$ (panel d).  
 With a further decrease in temperature, the resonances merge into one signal and follow together until the lowest temperature  of our measurements, 4.2\,K (see Fig.~\ref*{fig:H(T)_c}a). 

A resonance signal of irregular shape (consisting of two merged together components) cannot be described by a single line of Dysonian or Lorentzian shape.	
The spectrum  can be approximated only under the assumption that the two signals are superimposed on each other (see a note in caption to Fig.~\ref{fig:spectra}). 
This merging of resonance lines  is understandable, since in a weak perpendicular field an antiferromagnet behaves like a weak ferromagnet (where sublattice magnetization vectors undergo spin canting).
Other narrow signals noticeable on panels (c, d, e) in the resonance absorption spectra in weak fields $H<0.5$T do not depend on temperatures and are associated with impurities in the sapphire from which the resonator is made. 

The fact that the low-field M-line is observed predominantly in the geometry ${\bf H}\parallel (ab)$ and ${\bf H}_\sim \perp (ab)$ indicates that the resonating magnetic moments also lie in the $ab$ plane, and the corresponding anisotropy constant $K_u^{(M)}$ is large, much greater than  $K_u^{(AFM)} = 1.4\times 10^5$ J/m$^3$, determined in Ref.~\cite{golov_JMMM_2022} for  magnetic moments of Eu-planes belonging to the AFM lattice (see Fig.~\ref{fig:crystals}). The large value of $K_u$ is another indirect evidence of ferromagnetic origin of the M line in the resonance spectrum.

\subsubsection{Temperature dependence of the resonance field} 

\begin{figure*}
\includegraphics[height=172pt]{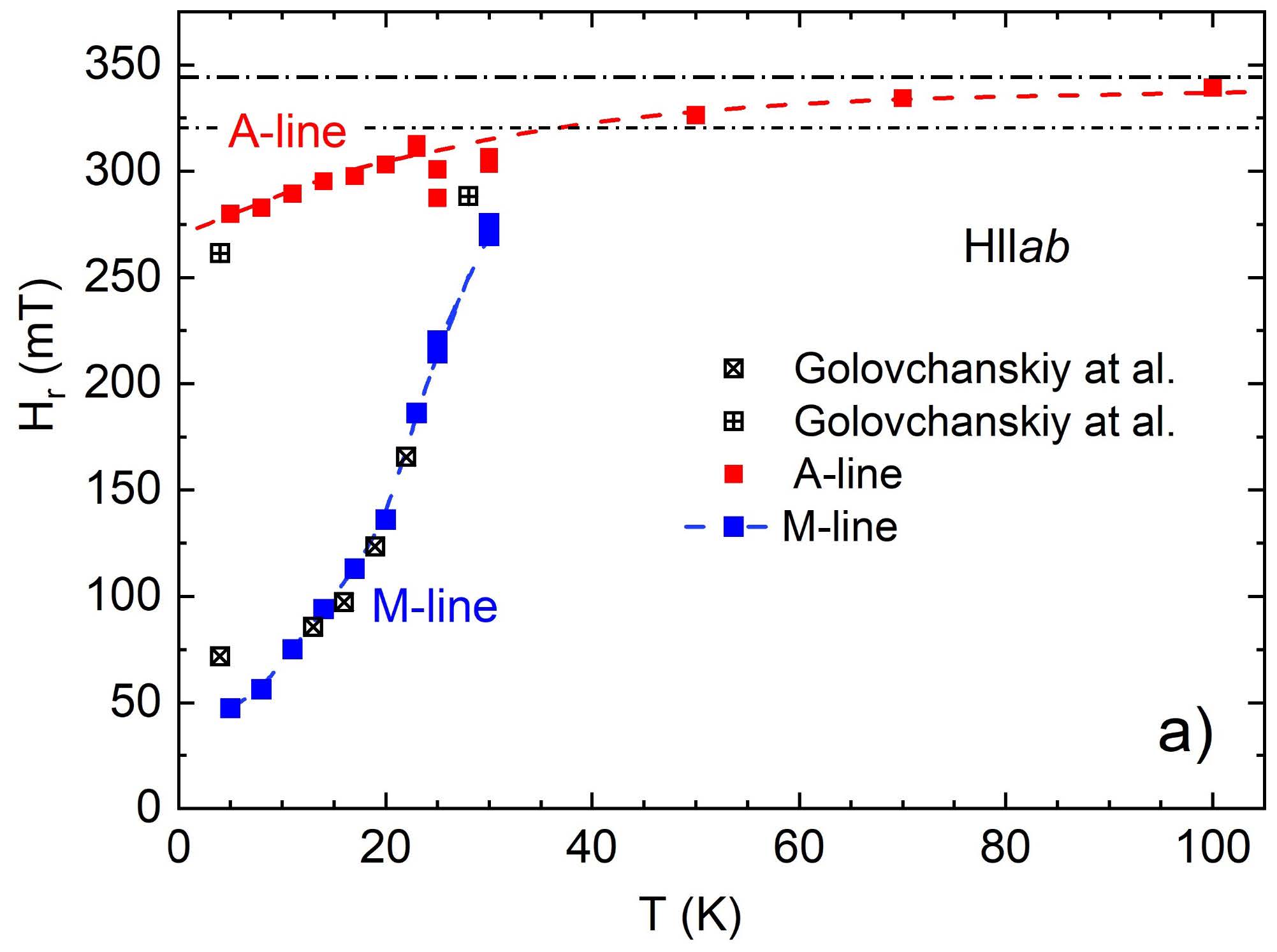}
\includegraphics[height=174pt]{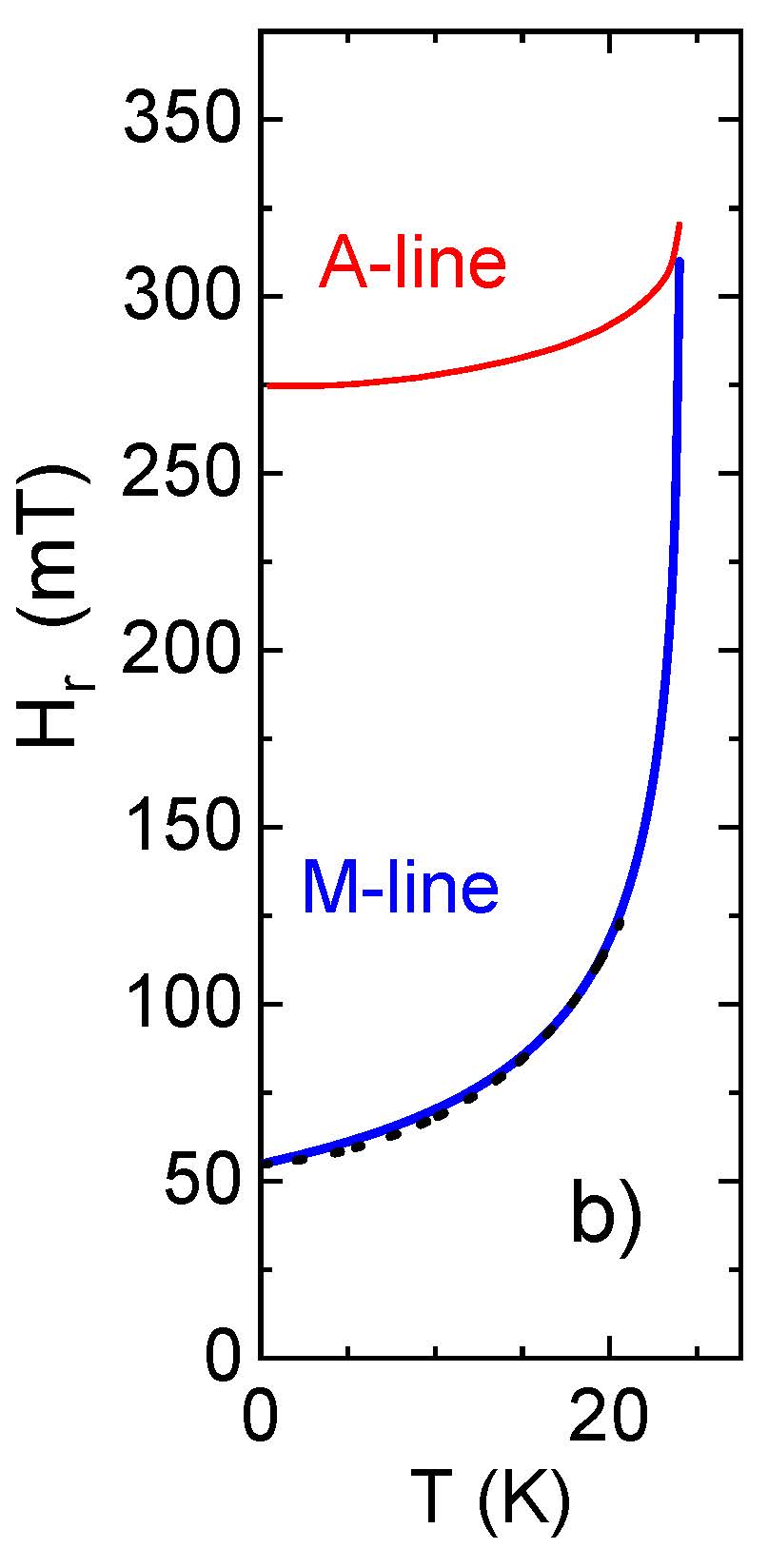}\\
\includegraphics[width=200pt]{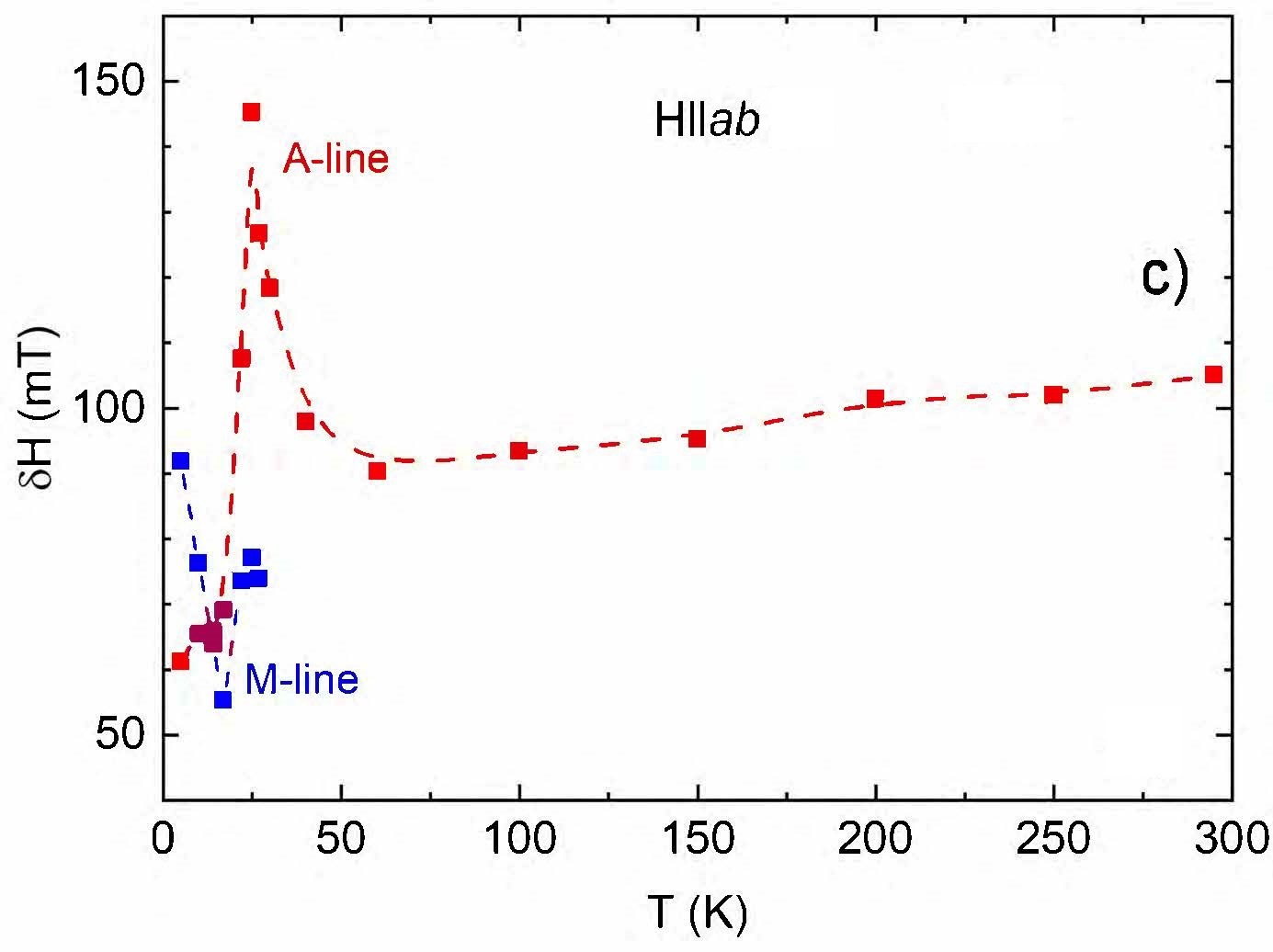}
\includegraphics[width=200pt]{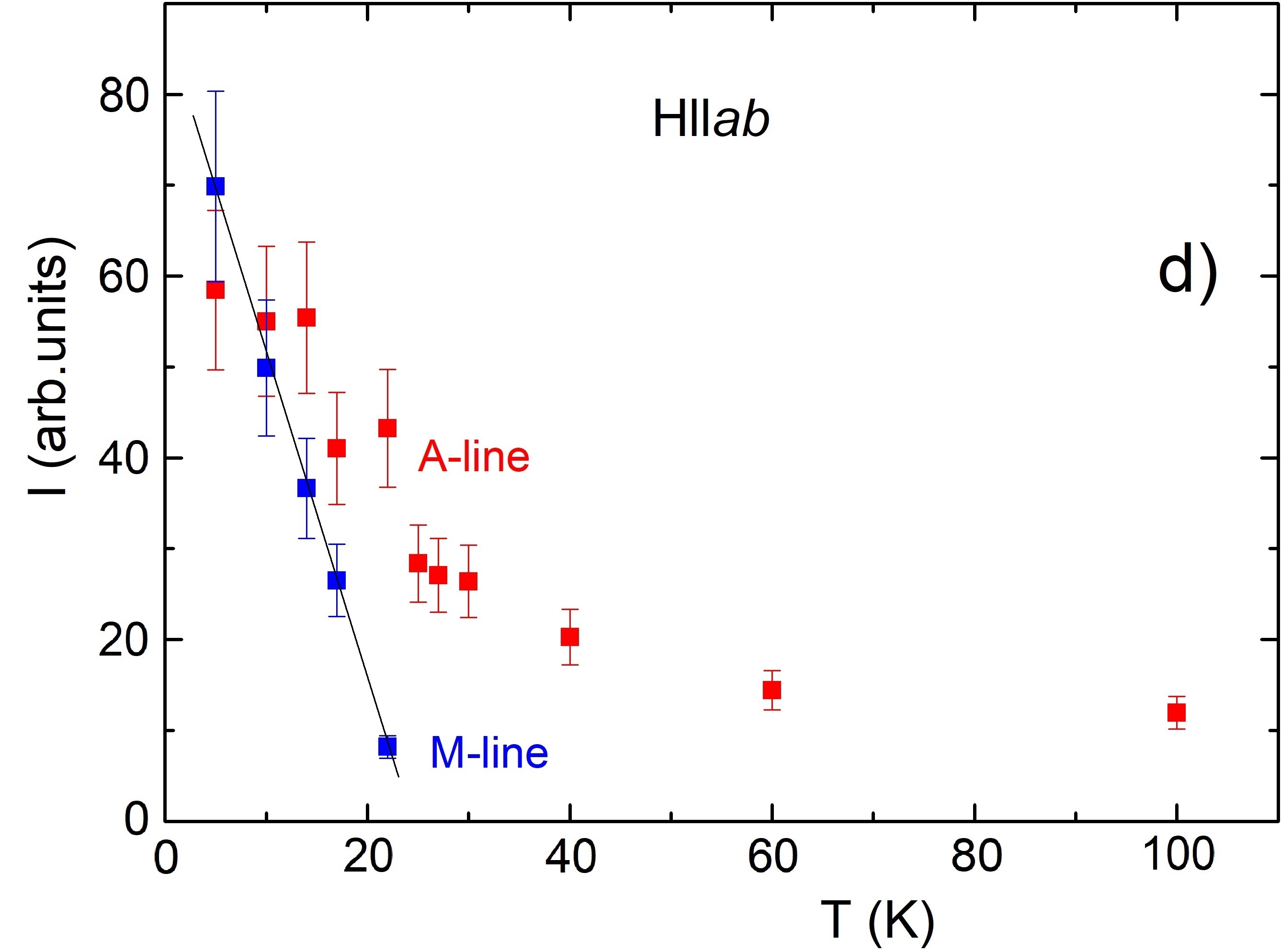}
	\caption{(a) Temperature dependence of the resonance field $H_r$\ for $H\|(ab)$. 
	Red solid squares are from our ESR measurements, crossed squares are data extracted from Ref.~\cite{golov_JMMM_2022} for the A-line.  Blue solid squares are the data of present ESR measurements at $f=9.735$\,GHz frequency, crossed squares depict results of \cite{golov_JMMM_2022} for the M-line. The dash-dotted lines correspond to $g=2.00$ (upper) and $g=2.155$ (lower);
	 (b) Calculated dependences $H_r(T)$ for resonance lines A and M. For A-line it is obtained using Eqs.~(1) \& (3);  for M-line: the solid curve  - using Eqs.~(4) \& (5); dotted line - using Eqs.~(4) \& (6). 
	 (c) Temperature dependence of the width for the A- and M- resonance lines.
(d) Temperature dependence of the integral intensity of two ESR signals, obtained by multiplying 
their amplitude by the square of the width. The straight line is a guide to the eyes.
}
	\label{fig:Hr(T)_ab}
\end{figure*}

\underline{ $\mathbf{H}\|(ab)$}, $\mathbf{H}_{\sim}\perp (ab)$. Figure \ref{fig:Hr(T)_ab}a shows the resulting temperature dependence of the resonance field in the geometry $\mathbf{H}\|(ab)$. Two resonances are present at low temperatures in the AFM region, in agreement with Ref.~\cite{golov_JMMM_2022}.  
The two lines tend to merge at temperature about 25-30K, close to the Neel temperature $T_N$. 
Along with our data, several points $H_r(T)$ from the work \cite{golov_JMMM_2022} are plotted in the same figure \ref{fig:Hr(T)_ab}a. One can see a good agreement between the positions of the observed resonances and the previously measured ones in Ref.~\cite{golov_JMMM_2022} on a similar crystal at variable frquency and in a constant field, at $T<24$K. From a comparison with the results of \cite{golov_JMMM_2022} we identify the ``upper'' branch in the figure \ref{fig:Hr(T)_ab}a  with the acoustic mode of the AFM resonance in the Eu spin system (``A-line'' \cite{golov_JMMM_2022}). 

The origin of the second, higher frequency mode, (i.e. the lower branch in Fig.~\ref{fig:Hr(T)_ab}a), 
was a mystery until now; an explanation of its origin is the main goal  of this work. 
We emphasize that  the two lines in the ESR spectrum are observed only in the AFM state ($T<24$K); 
at higher temperatures, both lines merge into one, in which the resonance field tends to a value close to the $g$-factor of paramagnetic Eu$^{2+}$ ions in the EuSn$_2$As$_2$ crystal ($g=2.005$) \cite{note-g-factor}. 

Figure \ref{fig:Hr(T)_ab}c shows temperature dependence of 
broadening of the resonance lines. 
The  width of the A line increases sharply as temperature decreases to 
$T=25$K, that is close to $T_N$.
Such behavior  is characteristic 
of AFM resonance and is associated with an increase in magnetization fluctuations in the vicinity of the transition.
On the contrary, the width of the M line with decreasing temperatures, after a slight decrease to 18\,K, 
increases down to the lowest temperature of our measurements $\sim 4$K, much lower than $T_N$.  
This behavior {\em is not typical} for AFM resonances, which confirms a completely different origin of the M line.

\begin{figure}
\includegraphics[width=220pt]{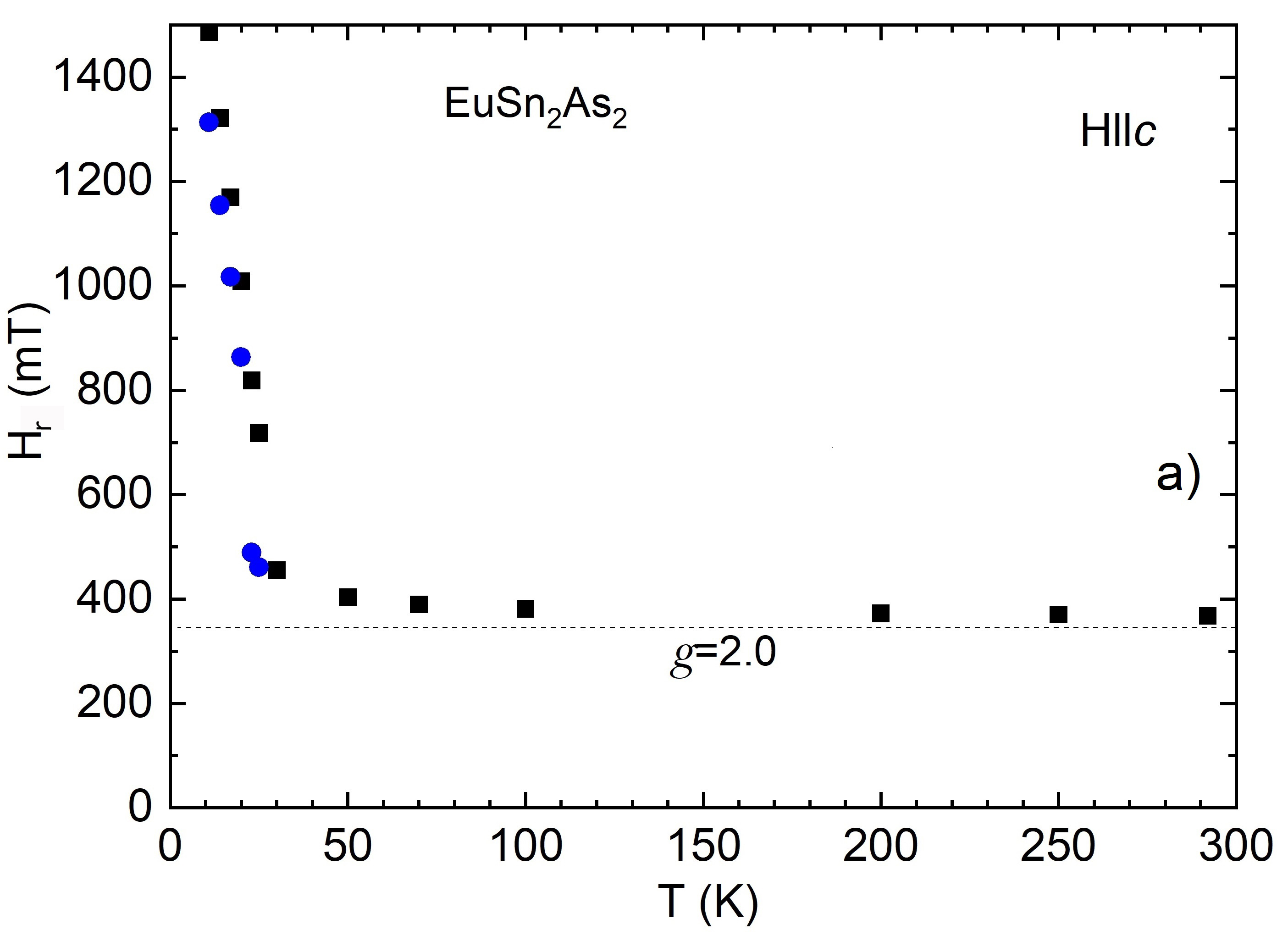}
\includegraphics[width=220pt]{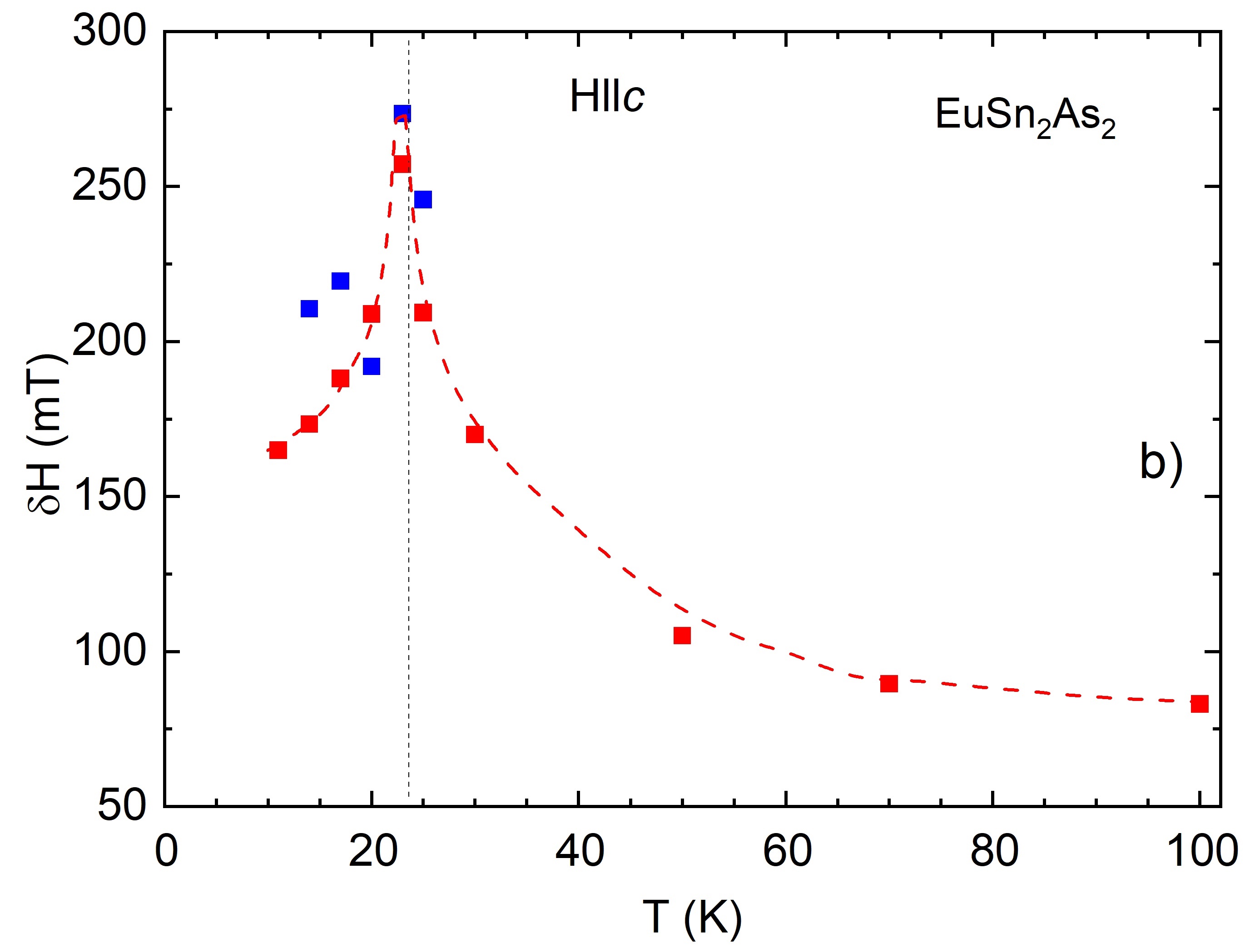}
\caption{Temperature dependence of (a) the resonance field $H_r$ for $H\|c$;  (b) the resonance signal width (peak to peak). Various symbols correspond to two lines in the resonance spectrum, for cases when they can be separated. $f=9.735$\,GHz.
	}
\label{fig:H(T)_c}
\end{figure}

\underline{$\mathbf{H}\|c$}.
Figure \ref{fig:H(T)_c}a shows the temperature dependence of the resonance field in the geometry 
$\mathbf{H}\|c$ and $\mathbf{H}_{\sim} \|(ab)$. 
As temperature decreases, in the vicinity of $T=T_N$ the resonance line 
splits into two components, which, with a further decrease in temperature, come closer together, 
and their resonance field increases sharply.  This behavior is qualitatively understandable for 
easy plane magnetization. It does not depend on how the spins in the $ab$ plane are ordered, FM or AFM. 
 
 The shift of resonance signals towards higher fields in both cases is determined by an increase 
 in the anisotropy field with decreasing temperature, because the anisotropy constant is negative, $K_U <0$, 
 (see \cite{gurevich}, Eqs.~(2.2.21) for ferromagnets and (4.3.2) for antiferromagnets). 
 The broadening of resonance signals near the magnetic ordering temperature, as in the parallel case, 
 is due to magnetization fluctuations, which affect resonant absorption in two ways:  
 on the one hand, fluctuating magnetic fields accelerate spin relaxation, causing uniform broadening. 
 On the other hand, the spread of local fields leads to inhomogeneous line broadening. 
 The resulting line is a curve that envelops uniformly broadened signals, each of which is shifted by the value of its local field.
  
\subsection{Comparison of measured spectra with model dependencies}

\subsubsection{Resonance spectra at constant temperature} 
Since our model of the two-phase state (AFM + FM) of the sample is in qualitative agreement with the experimental data, we will proceed to its quantitative comparison.
We first turn to the interpretation of the results of the $f_r(H, T=Const)$ measurements, which we performed earlier 
\cite{golov_JMMM_2022} on similar EuSn$_2$As$_2$ crystals, but using a broadband spectrometer.\\

\underline{Normal ``A-line'' of ESR.} 
For an ordinary acoustic-type AFM resonance at $T<T_N$ in weak fields $H \ll H_e$, the oscillation frequency is described by the linear relation (Eq.~(3) in \cite{macneil_PRL_2021} and Eq.~(3.40) in \cite{gurevich-melkov}): 
\beq
\omega_r=\gamma H_r\sqrt{1+\frac{M_{\rm eff}^0}{2H_e}}
\label{eq:A-line}
\eeq
Here $\gamma = g\frac{|e|}{2m_e c}$ is the  gyromagnetic ratio.
$g$ is the Lande $g$-factor, $e$ and $m_e$ are the free electron charge and mass,
$M_{\rm eff}^0$ - the effective saturation magnetization per unit volume in the  $T\rightarrow  0$ limit,  and $H_e$ is the interlayer exchange field. When calculating the model curve, we used the following  values: $ M_{\rm eff}^0\approx 1.3$\,T and $H_e \approx 1.8$\,T, determined in Ref.~\cite{golov_JMMM_2022} from magnetization field dependences for EuSn$_2$As$_2$ in the limit of $T\rightarrow 0$.

\begin{figure}
		\includegraphics[width=240pt]{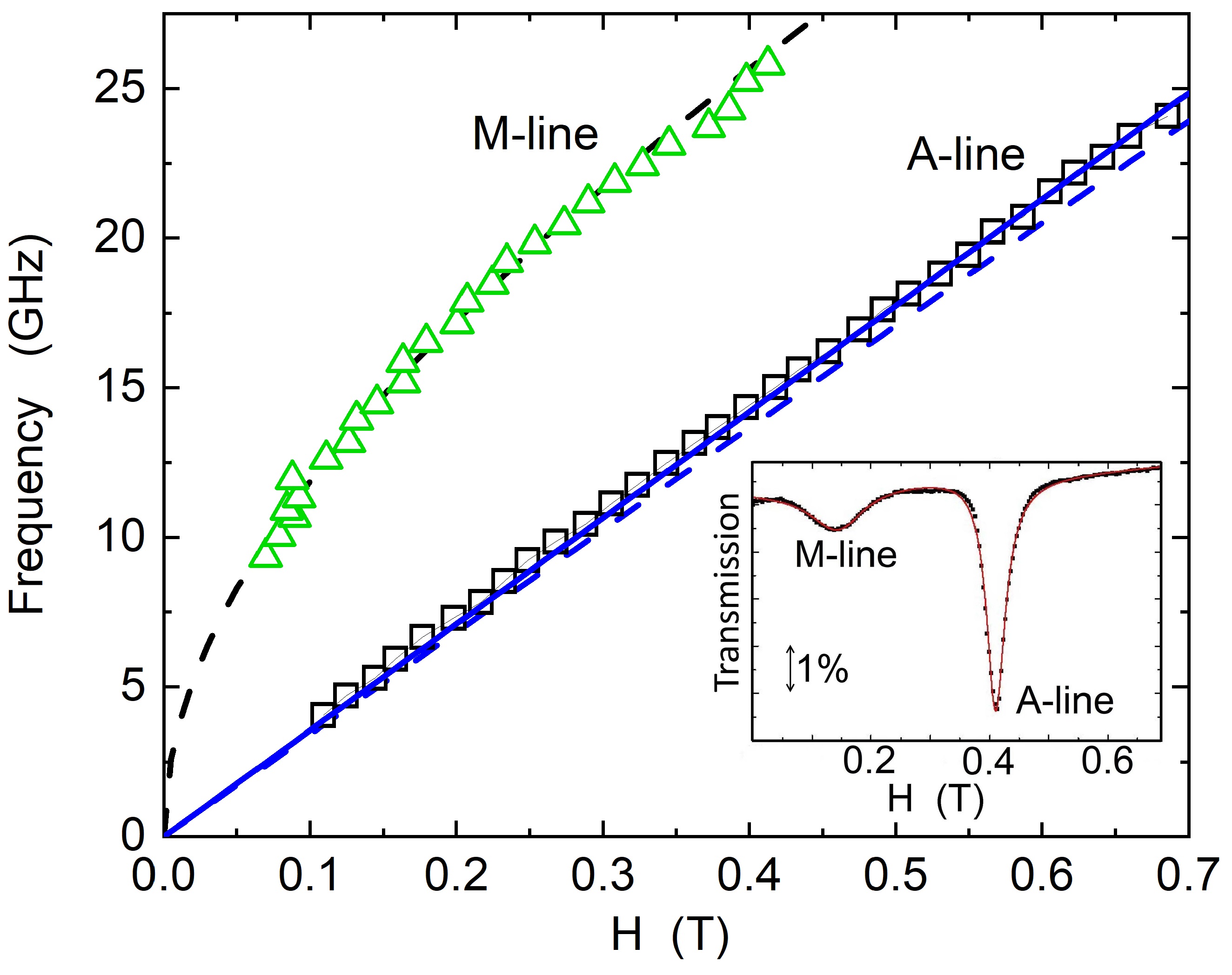}
	\caption{Experimental dependence of resonance frequency on magnetic field (symbols) 
at temperature $T=5$\,K (adapted from \cite{golov_JMMM_2022}), and their comparison with our model with no fitting parameters. The curves for A-line are calculated using Eq.~(\ref{eq:A-line}) with 
$g=2.09$ (solid line), and  with $g=2.14$ (dashed line).
The curve for M-line is calculated using  Eq.~(\ref{eq:M-line_f(H)}) with $M_{\rm def}^0=0.135$\,T. 
 The inset shows an example of the ESR spectrum measured at $T=4$\,K, and $f=15$\,GHz  
(adapted from \cite{golov_JMMM_2022}).
}
	\label{fig:simulation_f(H)}
\end{figure}

As Fig.~\ref{fig:simulation_f(H)} shows, with these parameters, Eq.~(\ref{eq:A-line}) well describes the isothermal linear dispersion (A-line) measured in \cite{golov_JMMM_2022}. 
We note that the experimental results $f_r(H)$\ are best described using value of $\gamma/2\pi =30.17$\,GHz/T, i.e. $g=2.155$. This value agrees  with  $g = 2.093$\ found from spectroscopic measurements of Zeeman splitting for the 7/2 spin state of Eu atoms \cite{furmann_JQS_2020}, 	as well as with the results of theoretical calculations \cite{smith_JOSA_1965, wyart_PhysScr_1985} ($g=2.094$\ and $g=2.082$).
However, it differs slightly from the value of 2.005 we obtained for ESR signals in the paramagnetic temperature range ($T>100$\,K). 
This difference in the $g$\ values, obtained from data at different temperature ranges ($T<T_N$\ and $T\gg T_N$), is possibly due to the influence of the environment 
in the crystal lattice, as well as to the simplified form of the used Hamiltonian, which takes into account only the 
interaction between the nearest neighboring Eu atoms.

\underline{Anomalous  ``M-line'' of ESR.}
In the work \cite{golov_JMMM_2022} it was shown that the anomalous high-frequency resonance line (M-line)  is not an optical mode of the AFM resonance \cite{gurevich-melkov}.
We interpret, therefore, the M-line as a {\em ferromagnetic} resonance originating from planar defects 
which have a non-zero magnetic moment \cite{ESA-defects}.  
In this case, we have: 
\beq
\omega^2 = \gamma^2 H[H+4\pi M_{\rm def}]
\label{eq:M-line_f(H)}
\eeq
We adopted here for the  field geometry in the plane of a thin plate, the components of the demagnetizing factor 
to be $N_x=N_z=0, N_y=4\pi$ (see \cite{kittel}, Eq. (17.49), and also \cite{golov-EuFeAs2_PRB_2022}).
In the above equation, $M_{\rm def}$ is the saturation magnetization created by ferromagnetic defects.
Dashed curve  in Fig.~\ref{fig:simulation_f(H)} demonstrates that 
Eq.~\ref{eq:M-line_f(H)} well approximates the experimental $f_r(H)$ dependence 
for the M-line, using a fixed $\gamma/2\pi= 29.96$\,GHz/T (i.e. $g=2.14$) value
and a single fitting parameter $M_{\rm def}^0=0.135$\,T. 
The value $M_{\rm def}^0$, obtained here from fitting, is further used as an estimate for the 
magnetization of the defect domain at $T\rightarrow 0$.

The value $M_{\rm def}^0= 0.135$T obtained from the approximation is by 9.6 times  less than the 
saturation value of the bulk magnetization of the ideal EuSn$_2$As$_2$  crystal in the AFM state: $M_{\rm eff}^0 =1.3$T \cite{golov_JMMM_2022}. This factor of 9.6 reduced $M_{\rm def}$ 
compared to $M_{\rm eff}$, is in a reasonable agreement with the discussed  above result of the DFT calculations \cite{ESA-defects}, according to which  only 1/7 Eu atoms participate in ferromagnetic ordering in the formula unit of the transition layer of the defect, Eu$_7$Sn$_{12}$As$_{14}$.

Indeed, according to this theoretical result, and also due to the fact that all magnetic moments 
(both AFM and FM) lie in the same easy $ab$ plane, it can be expected that the saturation magnetization of ferromagnetic moments should be  approximately an order of magnitude less than the saturation magnetization of an ideal antiferromagnetic crystal, $1.3$T \cite{golov_JMMM_2022}. Further reduction of the resonance signal amplitude may occure due to a non-zero share of  the sufficiently large defects; the latter ones may contain more than one domain (see Fig.~\ref{fig:defects}c),  behave as  local macroscopic antiferromagnets and don't participate in the FM-resonance at the 3cm-wavelength that is much larger than $L$, and $W$. 
We highlight, that low concentration of defects in the crystal ($\sim 3\%$) affects only the amplitude of the ESR signal, rather than its frequency.   

At such a low concentration of defects, the approximate equality of  amplitudes of the two signals, AFMR (A-line) and FMR (M-line), at first glance seems surprising. In addition, as we mentioned above, the saturation magnetization of ferromagnetic defects is almost 10 times smaller than that of  the antiferromagnetic volume. However, the reason for the approximate equality of the resonance amplitudes can be understood if we take into account that their integral intensity $I$ is proportional to the magnetization $I \propto M(T,H)/H$ \cite{kittel}. The saturation magnetization of the AFM volume is achieved in fields of the order of 4\,T \cite{golov_JMMM_2022}, whereas the field at which the AFM resonance is observed is a factor of 10 less, $H<0.3$\,T. 

Accordingly, in such low field, the AFM magnetization acquires
only 10\% of its saturation value, $M_{AFM}<0.1M_{eff}^0$, 
while  magnetization of the ferromagnetic regions has already reached the maximum value 
corresponding to the complete polarization of spins $M_{FM}\approx M_{def}^0$. Since $M_{def}^0 \approx 0.1M_{eff}^0$, the intensity of the two lines becomes approximately equal. 
This argument is confirmed by the temperature dependence of the integral intensities of the two signals, 
shown in Fig.~\ref{fig:Hr(T)_ab}d. One can see that as temperatre decreases the intensity of the A-line increases smoothly while the M-signal intensity begins to increase sharply  at $T=T_N$  and reaches values comparable to the A-signal below  $T\approx 15$\,K.

Good agreement between the model dependencies and isothermal spectra (Fig.~\ref{fig:simulation_f(H)}) confirms 
the proposed interpretation of the two ESR resonance lines 
in the AFM crystal as (i) the acoustic mode of AFM resonance in the bulk of the crystal (line A), and (ii) FM resonance on magnetic defects (line M). We use this interpretation and the model described above in the next section to compare it with the measured temperature dependencies of the ESR spectra. 

\subsubsection{Temperature dependence of the resonance field }
\underline{A-line. $\mathbf{H}\| (ab)$}\\
In the divalent state, Eu atom with the electron configuration $4f^7$ has a large and purely spin magnetic moment 
($J = S= 7/2$) with a value of $\approx 6.6\mu_B$, that is responsible for the magnetic 
properties of the Eu sublattice \cite{kim_PRB_2021}. 
Since the orbital moment of divalent Eu 4f is zero, the total magnetic 
 moment of Eu is fairly insensitive to the surrounding crystalline electric field.

We assume  that the major contribution to the temperature dependence of the resonance field in Eq.~(1) comes from
 the temperature dependence  of $M_{\rm eff}$, and neglect for simplicity by potential temperature variation of $H_e$. Then, with increasing temperature, the effective magnetization of individual FM layers $M_{\rm eff}$ in a EuSn$_2$As$_2$ crystal  in the AFM state decreases as \cite{gurevich-melkov, kittel}
 \beq
M_{\rm eff}(T) =M_{\rm eff}^0 B_J\left(\frac{M_{\rm eff} H}{N k_B T}  \right)
\label{eq:M(T)}
\eeq
where $B_J$ is the  Brillouin function for $J=7/2$, $M^0_{\rm eff}=\gamma\hbar J N$ - saturation magnetization (at $T\rightarrow 0$), $N$-  number of magnetic moments per volume. 

The temperature dependence $M_{\rm eff}(T)$, found by numerically solving the above nonlinear equation,
without adjustable parameters, after substituting it into the formula (\ref{eq:A-line}), provides a qualitative agreement with the measured temperature dependence 
of the resonance magnetic field $H_r(T)$ for A-line, as  Fig.~\ref{fig:Hr(T)_ab}b shows.
\vspace{6pt}

\underline{M-line. $\mathbf{H}\|(ab)$}\\ 
Choosing positive solution to the quadratic equation (\ref{eq:M-line_f(H)}) for FM resonance we have:
\beq
H_r =\frac{1}{2}\left(-4\pi M_{\rm def} +\sqrt{(4\pi M_{\rm def})^2 +4(\omega/\gamma)^2}  \right).
\label{eq:M-line_Hr(T)}
\eeq
Here $M_{\rm def}$ is the saturation value of ferromagnetic magnetization of planar defects per unit defect volume. 

In order to find temperature dependence of the resonance field for the M-line, 
we use the textbook equations applicable in the vicinity of $T_c \approx T_N$ \cite{Landau}:
\beq
M_{\rm def}(T)= M_{\rm def}^0\sqrt{(T_c-T)/T_c}
\label{eq:vicinity_Tc}
\eeq
and in the vicinity of $T=0$ we use the Bloch law \cite{ashcroft&marmin}
\beq
M_{\rm def}(T) = M_{\rm def}^0\left[1- \beta \left(\frac{T}{T_c}\right)^{3/2}\right]
\label{eq:Tzero}
\eeq
with the zero-field magnitude $4\pi M_{\rm def}^0=1.696$\,T which
was determined above by approximating the isothermal dependence of the M-line frequency on magnetic field (Fig.~\ref{fig:simulation_f(H)}).

The resulting model dependecies $H_r(T)$ are shown in Fig.~\ref{fig:Hr(T)_ab}b.
For comparison, we plot on the panel (b) two  $H_r(T)$ dependencies for the M-line which use (i) Eq.~(\ref{eq:vicinity_Tc}) with no fitting parameters (solid curve), and (ii) the low-temperature part of Eq.~(\ref{eq:Tzero}) (dotted curve) with prefactor $\beta$ adjusted to match the solid curve at higher temperatures; 
both curves almost conside. 

Since the ferromagnetic order emerges in the AFM state \cite{ESA-defects}, in this modelling, for simplicity, we assumed the same critical temperature (Curie) for the nanodefect as the Neel temperature of the bulk crystal, $T_c=T_N= 24$K. One can see that the model temperature dependence of the resonance field for the M-line calculated with these parameters is qualitatively similar to the measured temperature dependence of the M-line (compare panels (a) and (b) in Fig.~\ref{fig:Hr(T)_ab}).

\section{Discussion}

As can be seen from Fig.~\ref{fig:Hr(T)_ab}, the model temperature dependencies of both lines agree qualitatively 
with  the measured ones. However, the resonance field in the model curves increases with temperature raising at first more slowly, and near $T_N$ - faster than in the experimental dependence.
At temperatures above $T_N$, both resonance lines merge into the single paramagnetic resonance line. 
A weak increase in the measured ESR resonance field with temperature is quite expected and corresponds to the disordering of the spin magnetic moments of Eu atoms.

Some difference in the character of the growth of the resonance field $H_r$ with increasing temperature in the model dependencies, compared with that observed in the experiment (Fig.~\ref{fig:Hr(T)_ab}a,b), may be due to the oversimplified character of the model in which we assumed complete independence of two spin subsystems - ideal AFM and ideal FM, and also ignored potential domain structure of some large defects. In reality, the crystal under study is a metamaterial in which the AFM and FM ordering and the resonance modes interact. This is indicated by the proximity of the onset temperatures of FM- and AFM resonances, i.e. $T_c\approx T_N$.

It should be noted that  the A-type AFM ordered layered crystal CrSBr, that is isostructural 
  to $\rm EuSn_2As_2$ and composed of ferromagnetically-ordered layers of chromium, 
exhibits a very much different microwave spectrum including two AFM magnon modes \cite{Cho2023}.
This dissimilarity  is caused by much more complex magnetic hamiltonian in this material, including bi-axial single-ion anisotropy and inter-plane exchange.

In general, our results  demonstrate that the antiferromagnetically ordered and chemically pure EuSn$_2$As$_2$ crystal is, in fact, a nontrivial object - the natural magnetic metamaterial containing crystalline FM-inclusions oriented in the basal plane of the parent AFM-crystal. We believe that planar nanodefects in layered compounds contaning magnetic atoms are characteristic  not only of EuSn$_2$As$_2$, but may also be present in other compounds. 

This is indirectly evidenced by the anomalous increase in AC susceptibility 
with decreasing temperature $T\ll T_N$, which is often observed  in a weak DC field not only in EuSn$_2$As$_2$ \cite{ESA-defects}, but also in Ca$_{1-x}$Sr$_x$Co$_{2-y}$As$_2$ \cite{sangeetha_PRL_2017},  KCo$_2$Se$_2$ \cite{yang_PRB_2013}, and others. In this regard, it is worthnoting the recent discovery of ferromagnetism in a MnBi$_2$Te$_4$ crystal caused by point defects - displacement of Mn atoms \cite{fukushima_PRM_2024}. Though for the purpose of this paper, the orientation of the FM
moments and their domain structure were insignificant,  this issue, in principle,  might be clarified by applying scanning SQUID or NV-center microsopy at low temperatires, below $T_N$.

\section{Conclusion}

In this work, we carried out detailed studies of electron spin resonance on the spins of Eu$^{(2+)}$ ions in a layered EuSn$_2$As$_2$ crystal. 
Thanks to the use of a spectrometer with cavity resonator with high sensitivity 
we were able to measure the spin resonance  over a wide temperature range, 4 - 300\,K, covering 
both the AFM spin ordered phase of Eu ($T<24$K) and the paramagnetic phase ($T>T_N$).

 The splitting of the resonance absorption line is observed in the whole AFM temperature range, 
 confirming the previous results obtained at constant temperature and variable microwave frequency \cite{golov_JMMM_2022}. The resonance line at higher  field corresponds, in all respects, to the usual acoustic mode of the AFM resonance. Its dependence on the magnetic field  is described quantitatively and on temperature - qualitatively, with no fitting parameters when using the AFM saturation magnetization value $M_{\rm eff}=1.3$\,T, 
 measured experimentally.

The main result of this work is the experimantal proof that the anomalous M-line in the ESR spectrum at 
 a lower resonance field, is the ferromagnetic resonance  associated with the presence of a small number of planar defects in the crystal. These structural defects, their crystal structure and chemical formula were determined by transmission electron microscopy,  energy-dispersive X-ray spectroscopy, and electron energy loss spectroscopy on samples from the same butch, and their ferromagnetic moment was established from the DFT calculations and precise magnetic measurements \cite{ESA-defects}.

Specifically, in this work we have shown that the magnetic field- and temperature- dependencies of the 
two resonance lines  in the ESR spectrum are described within the 
model considering the rare FM-defects imbedded into the bulk AFM crystal, and using the 
single parameter $M_{\rm def}^0$ - the saturation magnetization of the FM-defect.
This FM magnetization value agrees with the estimate 
from DFT calculations, TEM measurements, and from direct magnetization hysteresis measurements \cite{ESA-defects}.
The consistenecy of the model and the data confirms the concept  treating
the high purity layered EuSn$_2$As$_2$ crystal in the AFM domain of external parameters as the two-phase
metamaterial.

\section{Acknowledgements}
IIG, DEZ, RBZ, and YuIT acknoledge Russian Science Foundation for the financial support via grant \# 21-72-20153-P.
KSP, VAV, AVS, and VMP acknowledge support of the LPI State assignment of the Ministry of Science and Higher Education of the Russian Federation (Project No. 0023-2019-0005). 
Synthesis, crystal  growth, lattice structure measurements, samples characterization, and magnetic measurements were done using  equipment of the LPI shared facility center.


\begin{thebibliography}{100}
\bibitem{note:line-shape}The signal shape on the panels Fig.~\ref{fig:spectra}(d,e) has inverse asymmetry (compared to the Dyson spectrum), 
that is, the low-field extremum has a smaller amplitude compared to the high-field one (as opposed to Dyson's). This shape cannot be described by either a Dyson, Lorentz, or Gaussian shape.
For this reason, we approximated the resonance line by two signals, which, as a result of superposition of components with different signs (the minimum of one is superimposed on the maximum of the other), give such an unusual shape (with inverse asymmetry). The parameters of these two signals, obtained from fitting, are plotted on the temperature  dependencies of the resonance field, Fig.~\ref{fig:H(T)_c}a.
\bibitem{note-g-factor}
At high temperatures (significantly above $T_N$) the Lande $g$-factor is $g=2.005$ at $H\|(ab)$ and $g=1.92$ at $H\|c$ (as noted above). At temperatures around $T_N$ and below, one cannot introduce the $g$ factor. There, the position of the signal is determined by internal fields associated with exchange, anisotropy, etc.
\end{thebibliography}
\end{document}